\documentclass{article}
\usepackage[T1]{fontenc} 
\usepackage[utf8]{inputenc} 
\usepackage{ismir,amsmath,cite,url}
\usepackage{amsfonts}
\usepackage{graphicx}
\usepackage{color,booktabs}
\newcommand\myuparrow{\mathord{\uparrow}}

\usepackage{lineno}

\title{Hierarchical Timbre-painting and Articulation Generation}

\twoauthors
  {Michael Michelashvili} {Tel Aviv University \\ {\tt mosheman5@gmail.com}}
  {Lior Wolf} {Tel Aviv University\\ {\tt wolf@cs.tau.ac.il}}

\begin{document}

\maketitle
\begin{abstract}
We present a fast and high-fidelity method for music generation, based on specified f0 and loudness, such that the synthesized audio mimics the timbre and articulation of a target instrument. The generation process consists of learned source-filtering networks, which reconstruct the signal at increasing resolutions. The model optimizes a multi-resolution spectral loss as the reconstruction loss, an adversarial loss to make the audio sound more realistic, and a perceptual f0 loss to align the output to the desired input pitch contour. The proposed architecture enables high-quality fitting of an instrument, given a sample that can be as short as a few minutes, and the method demonstrates state-of-the-art timbre transfer capabilities. Code and audio samples are shared at https://github.com/mosheman5/timbre\_painting.
\end{abstract}
\section{Introduction}\label{sec:introduction}

The melody, as depicted by a sequence of notes, or alternatively by a sequence of frequencies, is one generic aspect of the musical experience. The dynamic loudness signal is another prominent aspect that is also almost instrument-invariant. Due to the invariance property of these two aspects, it is natural to employ them as specifications to the instrument-independent essence of a musical piece.

A prominent aspect that does depend on the instrument is the timbre. The music-AI task of timbre-transfer considers methods that receive, as input, an audio segment and a target instrument, and output the analog (melody preserving) audio in the target domain, by replacing the timbre of the original audio clip with that of the specified instrument.

{Another aspect that defines a musical instrument is articulation, or the joining-up of notes. Timbre transfer methods address this implicitly with varying degrees of success.} The physical properties of the instrument lead to constraints and subsequently different characteristic ways to move from one note to the next in a smooth manner. This aspect, therefore, varies considerably, e.g., between violin, guitar, and trumpet. 

\begin{figure*}[t]
 \centering
 \begin{tabular}{ccc}
 \includegraphics[width=.3\linewidth]{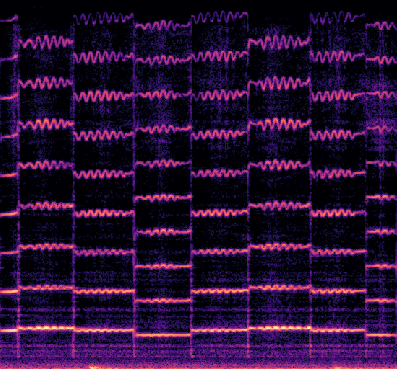}&
 \includegraphics[width=.3\linewidth]{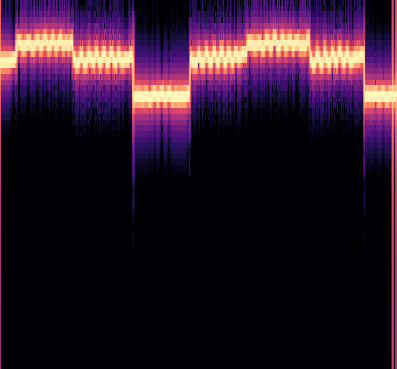}&
 \includegraphics[width=.3\linewidth]{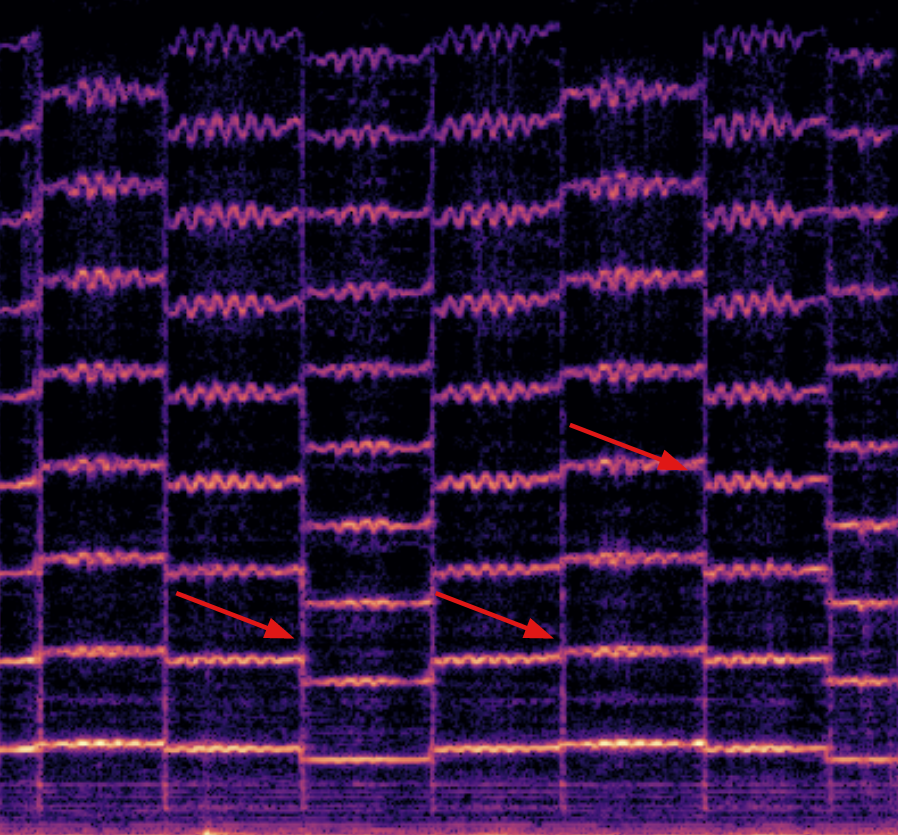}\\
 (a)&(b)&(c)\\
 \end{tabular}
 \caption{An illustration of our method's articulation capabilities. (a) The spectrogram of a violin audio. (b) The extracted fundamental frequency (f0) of the violin. 
 (c) The results of our method. Both the timbre and the articulations were manipulated. See arrows for a few specific locations where the violin's articulation is demonstrated.}
 \label{fig:articulation}
\end{figure*}

While this interpolation process is second nature for trained musicians, it can be sophisticated and involves the introduction of new frequencies that are not part of the original notes. See Fig.~\ref{fig:articulation}.

\begin{figure*}[t]
 \centering
 \begin{tabular}{ccc}
 \includegraphics[width=.3\linewidth]{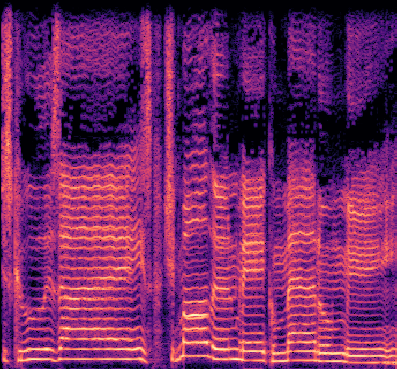}&
 \includegraphics[width=.3\linewidth]{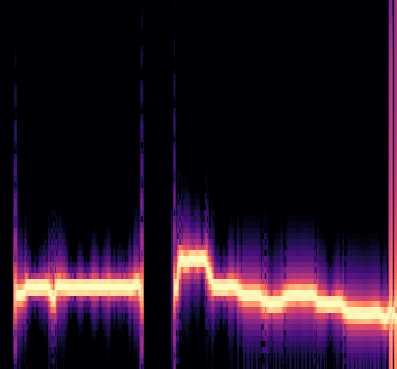}&
 \includegraphics[width=.3\linewidth]{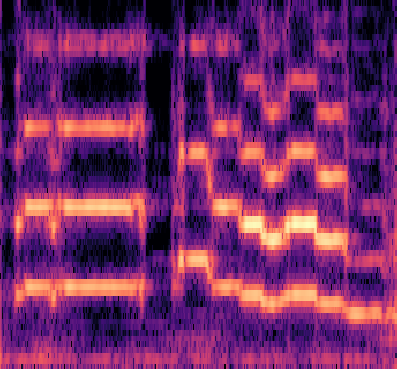}\\
 (a)&(b)&(c)\\
 \includegraphics[width=.3\linewidth]{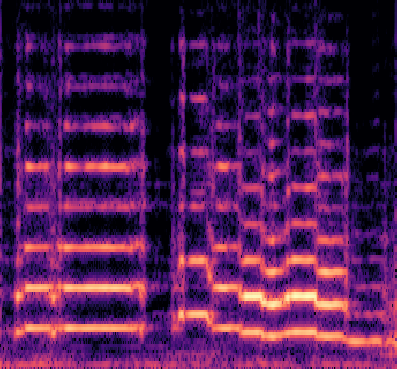}&
 \includegraphics[width=.3\linewidth]{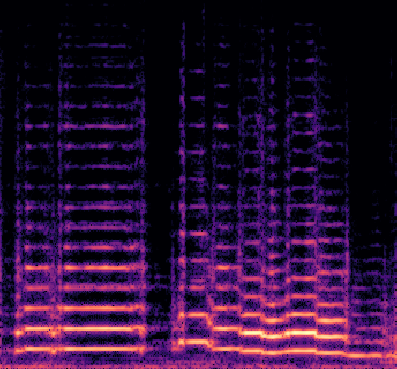}&
 \includegraphics[width=.3\linewidth]{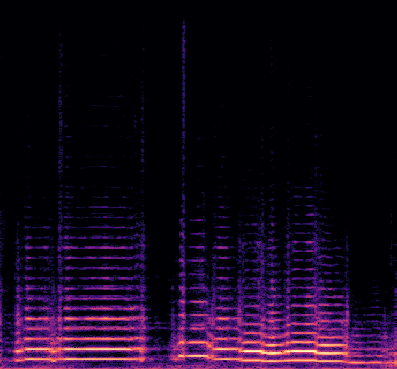}\\
 (d)&(e)&(f)\\
 \end{tabular}
 \caption{An illustration of the hierarchical generation process. (a) A spectrogram depicting the original melody as sang by a male singer. (b) The extracted fundamental frequency (f0) of the melody. (c-f) The generated audio for a saxophone from the coarsest scale $\bm{x}^0$ to the finest $\bm{x}^3$, respectively. While articulation-based manipulations are already seen in $\bm{x}^0$, the full effect of the timbre and spectral-quality is only observed at the final output $\bm{x}^3$.}
 \label{fig:scales}
\end{figure*}

\smallskip
In this work, we build a hierarchical music generator network. Given a fundamental frequency (f0) and loudness inputs, the network generates audio in four different scales. While the different scales share the same architecture, they have different roles. The first (lower) scale introduce the articulation, while the top scales introduce much of the timbre and the final audio-spectrum quality, which we call timbre-painting. See Fig.~\ref{fig:scales}. 

The model is trained on a relatively short sample from the target instrument, typically consisting of few minutes. The network is trained to minimize multiple losses: an adversarial loss encourages the output to be indistinguishable from audio in the output domain, multi-scale reconstruction losses in the frequency domain are used to ensure that the network can recreate the training sample, and the f0 of the output is compared to the specifications.

One possible application of the network is for the task of music domain transfer, similar to the application of other timbre-transfer methods. In this case, the f0 and loudness inputs are extracted from an existing audio clip and the network generates the analog music in the target domain. Our experiments show that our method generates audio that sounds more realistic and  is perceived to be of a better fit to the original melody than the recent state-of-the-art method DDSP~\cite{engel2020ddsp}.

\section{Related work}

The task of timbre-transfer was tackled by~\cite{timbretron}. An image-to-image pipeline that uses cycle consistency losses~\cite{CycleGAN2017} is applied to the audio domain by representing audio signals as 2D images with the Constant-Q-Transform (CQT). To move back from the CQT representation, a WaveNet~\cite{wavenet} synthesizer that is conditioned on CQT representation was used. Another prominent work~\cite{mor2018autoencoderbased} suggested to learn the audio melody by using a WaveNet Autoencoder architecture~\cite{nsynth2017}. One ``universal'' encoder is used to represent melody from raw data, and multiple domain-specific decoders are used for audio generation. By presenting domain-adversarial loss on the encoding, this method represents only the domain-invariant data needed for generation, which is predominantly the melody. Even though this method presents impressive results on timbre transfer and audio translation, it has few major disadvantages: the reliance on large amounts of data, and the heavy computation resources required (tens of GPUs).

The differentiable digital signal processing (DDSP) method~\cite{engel2020ddsp}, which was proposed recently, is much more efficient with regards to both data and computational needs. The method presents a DSP hybrid model in which a synthesizer with learned parameters is used. Like our method, DDSP conditions the signal generation on f0 and the loudness signal. Therefore, it can apply timbre-transfer to any audio for which a pitch tracker, e.g., CREPE~\cite{crepe}, can successfully extract the f0 signal.

DDSP and other methods~\cite{nsf2020music} follow the high fidelity speech synthesizer of~\cite{wang2019neural} in employing convolutional neural networks as shape-shifting filters to a sine-wave input. While many speech generation techniques condition the network on the f0 signal, this line of methods employ the corresponding sine-wave. 

Hierarchical generation was shown to be effective for image generation tasks. The progressive GAN method~\cite{karras2017progressive} breaks down the generation scheme into cascading generators and discriminators,  improving the image generation quality and stabilizing the training process. The SinGAN method~\cite{singan2019} performs convincing image-retargeting and image generation, using multi-scales learning from a single input image.

\section{Method}\label{sec:method}

Our method is hierarchical and consists of generators in four different scales. All generators have the same architecture of a non-autoregressive WaveNet applied on (scale-dependent) input and conditioned on extracted audio features on each scale.  The learning process is optimized to: (i) decrease the distance between the spectral representations of the generated and the target audio, (ii) minimize pitch perceptual loss in order to improve pitch coherence, and (iii) create realistically sounding examples by the usage of an adversarial loss.

\subsection{Input Features}

An audio sample is denoted by $\bm{x^n} = (x^{n}_1,\ldots, x^{n}_T)$, where T is the length of the signal and $n$ is the finest scale we consider. The scaled version of it are denoted by $\bm{x^{n-1}}$, $\bm{x^{n-2}}$, up to $\bm{x^0}$, which is the coarsest scale. The scaling is carried out by down-sampling, 
\begin{equation}
    x^{n-1}[t] = \sum_{k=0}^{K-1}{x^{n}[tM-k]h[k]}
\end{equation}
Where $M$ is the reduction factor, $h$ a FIR anti-aliasing filter and K the length of the filter. 

In our experiments, we use four scales $j=0..3$. The finest generates audio in 16 kHz, while the coarsest generates audio in 2 kHz. {We chose the coarsest scale to be as small as possible on the articulation generation phase, yet to include the f0 signal of our target instruments (max of 1kHz as given by Nyquist rule)}

In our method, audio is generated based on the specifications of the loudness of the output audio and its pitch. The other characteristics (timbre, articulation, and spectral quality) are being added by the model, based on the training sample. The loudness is given, following~\cite{loudness}, by the A-Weighting scheme, which is a weighted sum of the log of the power spectrum. We denote the loudness extraction computation by $\text{loud}(\bm{x^j})$, which is a 1D signal of a length that is 32 times shorter than the length of the input $\bm{x^j}$, $j=0..3$, due to the power spectrum extraction.

The fundamental frequency f0, which is also a 1D signal, is extracted using the CREPE pitch tracking network~\cite{crepe}, as is done in~\cite{musicwavernn ,engel2020ddsp}. We denote the extracted signal by $f0(\bm{x^n})$ and compute it only at the finest-resolution scale. The CREPE network has a resolution of 250Hz, which differs from the sampling rate of our network. However, this conditioning is provided as a sine-wave at the resolution of the coarsest layer (2kHz).

Specifically, following previous work in speech~\cite{wang2019neural} and music synthesis~\cite{engel2020ddsp, nsf2020music}, we apply what is known as ``neural source-filtering''. In this technique, instead of conditioning the generated sample directly on the extracted f0 signal, the generator is conditioned on a raw waveform that is synthesized via a single sinusoid sine-excitation, calculated from upsampled $f0(\bm{x^n})$. The $f0$ is downsampled by 32 from the input signal $\bm{x}^n$ and the coarsest scales $j=0$, which is generated first, has a frequency that is one eighth of the original audio. Scaling is, therefore, by a factor of 4. We denote the generated waveform by $\eta(f0(\bm{x^n}))$. \begin{equation}
    \eta(f0(\bm{x^n})) = sin(\sum_{k=0}^{T}2\pi{\myuparrow f0(\bm{x^n})_k/f_s)},
\end{equation}
where $f_s$ is the sample rate of the audio and $\uparrow$ denotes an upsampling operator.
 
\subsection{Hierarchical Generation}

The generated waveform $\eta(f0(\bm{x^n}))$ serves as the input to the lowest scale generator in the hierarchy, which is denoted by $G^0$. Similarly to our other generators and unlike conventional GAN generators, the generator does not receive random noise as input. 

In our method, we propose a conceptual relaxation to the audio generation task, and divide the generation into two distinct phases: timbre painting and articulation on the lowest scale, followed by upsampling networks which learn to generate higher resolution audio based on the previous scale. By doing so, we separate what we consider the most difficult part in the generation, namely converting a sine wave into well-articulated music, from the aspects of timbre painting and spectral quality adjustment. Therefore, fewer errors are introduced during the generation process and the method produces more coherent audio samples. 

Denote by $z^j=\text{loud}(\bm{x}^j)$. 
A set of input encoding networks $E^j$ transforms the raw input signal $z^j$ into a sequence of vectors, which $G$ is conditioned upon.

The lowest scale generator operates as follows:
\begin{equation}
    \hat{\bm{x}_0} = G^0(\eta(f0(\bm{x}^n)), E^0(z^0))\, , \label{eq:g0}
\end{equation}
where the second input is the conditioning signal.

\begin{figure*}
 \centerline{
 \includegraphics[width=\linewidth]{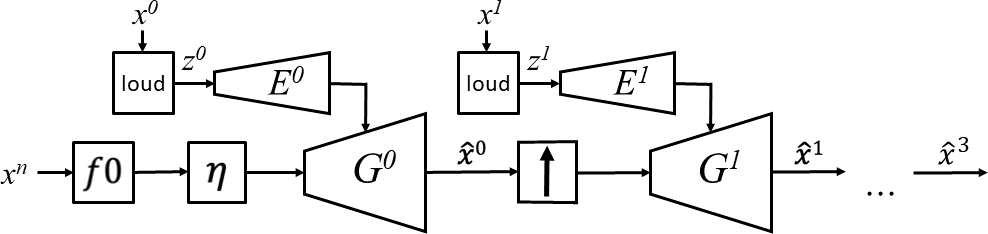}}
 \caption{An illustration of the generation process. The generator of the coarsest scale receives as input a sine-wave that is based on the fundamental frequency of the input sample. All generators are conditioned on the loudness signal of the appropriate scale. The output of the first generator $G^0$ serves as the input for the subsequent generator $G^1$ and so on.}
 \label{fig:arch}
\end{figure*}

The following generation steps receive as input the output of the previous scale generator:
\begin{equation}
    \hat{\bm{x}}^j = G^n(\myuparrow(\bm{\hat{x}^{j-1}}), z^j), \label{eq:gn}
\end{equation}
where $\myuparrow(\bm{\hat{x}_{n-1}})$ is an upsampled signal that matches the next scale. An illustration of the generation process is given in Fig.~\ref{fig:arch}.

\subsubsection{Architecture}

The architecture of the generators and discriminators is similar to that of\cite{parallelwavegan}. Each generator is composed of 30 layers stacked into three stacks. The kernel size is 3, using 64 residual channels and 64 skip channels. The dilation is exponentially growing in each stack, providing a receptive field of 3072 samples, which translates to a window size of 1.5sec on the lowest scale and 192ms on the finest.

The input encoder $E^j$ is composed of instance normalization, followed by 1D-convolution with kernel size of 1 that is applied on the condition input $z$. The number of output channels is 80. 
The output of $E^j$ is provided after upsampling via convolutional layers and nearest neighbor interpolation to the temporal dimension of the input signal.

Training involves a set of discriminators $D^j$, one per scale. Each discriminator is composed of 10 layers of 1D-convolution, followed by leakyReLU with negative slope of 0.2. The kernel size is 3, and 64 channels are used per layer. The dilation is growing linearly. Weight normalization is applied both on the generator and the discriminator.

\subsection{Training}

The learning setup and objective functions are the same for all the scales, with respect to the target audio signal. Conveniently, each generator $G^j$ is trained separately, after the previous generator $G^{j-1}$ is completely trained. We found that using the weights of the previous scale generator $G^{j-1}$ to initialize the weights of $G^j$ leads to faster convergence than random initialization on every scale. Similarly, the discriminator $D^j$ that provides the adversarial training signal to the generator $G^j$ is initialized based on $D^{j-1}$.

At each scale $j$, we obtain a training set $S^j$ by dividing the training sample, after it has been downsampled to scale $j$ to audio clips $\bm{x}^j$ of length 2sec. 

\subsubsection{Objective function} A time-frequency reconstruction loss is used to align to the generated audio sample with the target audio. Specifically, the spectral amplitude distance loss~\cite{oord2017parallel, arik2018fast}, in multiple FFT resolutions~\cite{wang2019neural, parallelwavegan, engel2020ddsp} is used. For a given FFT size $m$, the spectral amplitude distance loss is defined as follows: 
\begin{align}
    \mathcal{L}_{\text{recon}}^{(m,j)} = & \sum_{\bm{x}^j \in S^j} \left( \frac{\| |\text{STFT}(\bm{x}^j)| - |\text{STFT}(\hat{\bm{x}}^j)| \|_F}{\| \text{STFT}(\bm{x}^j) \|_F} \right. \nonumber\\ & + \left. \frac{\| \log|\text{STFT}(\bm{x}^j)| - \log|\text{STFT}(\hat{\bm{x}}^j)| \|_1}{N} \right)
\end{align}
where $\hat{\bm{x}}^j$ is given by Eq.~\ref{eq:g0} and Eq~\ref{eq:gn},  $\| \cdot \|_F$ and $\| \cdot \|_1$ denotes the Frobenius and the $L_1$ norms, respectively.  {The first element in the sum penalizes dominant bins in the magnitude while the second penalizes the silent parts.} $\text{STFT}$ denotes the {magnitude of a} Short-time Fourier transform with $N$ elements in the spectrogram. 

The multi-resolution loss is defined as the mean of the above loss for multiple scales:
\begin{equation}
    \mathcal{L}_{recon}^j =  \frac{1}{N_M}\sum_{m \in M} \mathcal{L}_{recon}^{(m,j)}
\end{equation}
where $M = [2048, 1024, 512, 256, 128, 64]$ and $N_M=6$ is the number of FFT scales. {Using the multi-resolution loss, we implicitly constrain the phase of the output signal to be correct and prevent artifact noises.}

To make the generated quality of the audio signals sound realistic, we introduce an adversarial loss. On each scale, we apply a different discriminator $D^j$ to account for different statistics between scales. 
We follow the least-squares GAN\cite{mao2017least}, where the discriminator minimizes the loss
\begin{equation}
\label{eq:discriminator}
\mathcal{L}_D^j = \sum_{\bm{x} \in S^j}{[|| 1 - D^j(\bm{x}^j) ||_2^2 + || D^j(\hat{\bm{x}}^j) ||_2^2]}\\
\end{equation}

Each trained generator $G^j$ minimizes the adversarial loss (recall that $\hat{\bm{x}}^j$ is computed with $G^j$):
\begin{equation}
\mathcal{L}_{adv}^j = \sum_{\bm{x}^j \in S^j}|| 1 - D^j(\hat{\bm{x}}^j) ||_2^2
\end{equation}

To further improve the generation quality, we add a perceptual loss~\cite{Johnson2016Perceptual} on the generator output, using the CREPE network~\cite{crepe}.  Denoting the mapping between the input signal $\bm{x}$ and the intermediate activations the CREPE network as $h(\myuparrow\bm{x})$, which requires an upsampling to 16kHz, this loss takes the form:
\begin{equation}
\mathcal{L}_{percep}^j = \sum_{\bm{x}^j \in S^j} \|h(\myuparrow\bm{x}^j) -h(\myuparrow\hat{\bm{x}}^j) \|_1\,.
\end{equation}
The optimization with this loss requires the upsampling operator to be differentiable.

In order to support a more direct comparison of the methods, following DDSP~\cite{engel2020ddsp}, the fifth max-pool layer of the small CREPE model is employed. 

Overall, the optimization loss for a generator $G^j$, is defined as: 
\begin{equation}
\label{eq:generator}
\mathcal{L}^j_G =  \mathcal{L}_{recon}^j + \alpha \mathcal{L}_{adv}^j  + \beta \mathcal{L}_{percep}^j
\end{equation}
where $\alpha, \beta$ are weight factors that balance the contribution of each loss term.

\section{Experiments}\label{sec:experiments}

\begin{table*}[t]
 \centering
 \begin{tabular}{lcccc}
  \toprule
  & \multicolumn{2}{c}{Target Similarity} & \multicolumn{2}{c}{Melody Similarity} \\
  \cmidrule(lr){2-3}\cmidrule(lr){4-5}
   Instrument/Method & DDSP & Our & DDSP & Our\\
  \midrule
  Cello & 4.11 $\pm$ 0.16 & 4.24 $\pm$ 0.16 & 4.00 $\pm$ 0.32 & 4.01 $\pm$ 0.49\\
  Saxophone & 3.09 $\pm$ 0.53 & 3.47 $\pm$ 0.54 & 3.87 $\pm$ 0.41 & 3.91 $\pm$ 0.53 \\
  Trumpet & 3.29 $\pm$ 0.45 & 4.01 $\pm$ 0.33 & 3.99 $\pm$ 0.29 & 4.11 $\pm$ 0.51 \\
  Violin & 4.02 $\pm$ 0.35 & 4.13 $\pm$ 0.27 & 4.13 $\pm$ 0.39 & 4.22 $\pm$ 0.39 \\
  \midrule
  All samples & 3.63 $\pm$ 0.60 & 3.96 $\pm$ 0.46 & 4.00 $\pm$ 0.36 & 4.06 $\pm$ 0.50\\

  \bottomrule
 \end{tabular}
 \caption{MOS evaluation for the timbre transfer task for multiple target instruments.}
 \label{tab:1}
\end{table*}

We conduct timbre-transfer experiments for multiple instruments, and compare the results to the state-of-the-art timbre transfer method DDSP~\cite{engel2020ddsp}.
 
\subsection{Datasets}

 We used the University of Rochester Music Performance (URMP) dataset~\cite{urmp}, a multi-modal audio-visual dataset containing  classical music pieces. The music is assembled from separately recorded audio stems of various monophonic instruments. For our experiments, we used only the separated audio stems for each instrument. f0 extraction was carried out by CREPE\cite{crepe}, although the URMP dataset provides ground truth melody signals, since we wanted to apply similar methods during train and test. 
 
 We trained both the baseline DDSP~\cite{engel2020ddsp} method and our model on generating four different instruments from the URMP dataset: cello, saxophone, trumpet and violin. As a prerocessing step the audio files were resampled to 16kHz. To improve the ability of learning meaningful f0 representation we removed in each dataset samples which achieved less than 0.85 mean confidence on CREPE extractor. Each dataset was separated into a training and evaluation set by 0.85/0.15 split. After the preprocessing, we ended up with small dataset sizes: 6.5 minutes of cello, 6 minutes of saxophone, 17 minutes of trumpet and 39 minutes of violin.

\subsection{Experiment Setup}

Our models were trained with $\alpha$=1 and $\beta$=1. We used the Adam optimizer~\cite{adam} with a learning rate of 0.0005 for the generators and 0.0001 for the discriminators. Each scale was trained for 120K iterations, with batch sizes of 32, 16, 8 and 4, from coarsest to finest. The learning rates were halved after 60K iterations. The discriminators were introduced to the training process on iteration 30K. 
To improve the robustness of our method we added a random Gaussian noise with a standard deviation of 0.003 to the $\eta(f0(\bm{x^n}))$ signal, inspired by~\cite{wang2019neural}. 

For the baseline evaluation of the DDSP method, the open source GitHub implementation\footnote{https://github.com/magenta/ddsp} provided by the authors of~\cite{engel2020ddsp} was used. The  experiments were carried out for 100K iterations with a batch size of 16. The hyper-parameters used are the ones provided by the recipe available in that repository. 

\subsection{User Study}

To inspect the results of the timbre transfer experiments we carried out a mean opinion scores (MOS) evaluation. We sampled six audio clips varying from 5-10s, long enough for good evaluation. The origin instruments are: clarinet, saxophone, female singer, male singer, trumpet and violin. For each audio sample, we conducted timbre transfer using the four models of the target instruments, resulting in a matrix of 24 inspection files for our method and 24 for the baseline. The timbre-transfer was done by extracting the loudness and pitch features from the source audio, aligning pitch key to the target (if needed) and generation procedure. The evaluations samples are available in the supplementary material. Twenty raters were asked to rate the generated outputs by two criteria: (i) target similarity to the transferred instrument, and (ii) the melody similarity to the original tune. Scores vary on a scale of one to five.

\subsection{Results}

\begin{figure}[t]
 \centering
 \begin{tabular}{c}
 \includegraphics[width=.89263\linewidth]{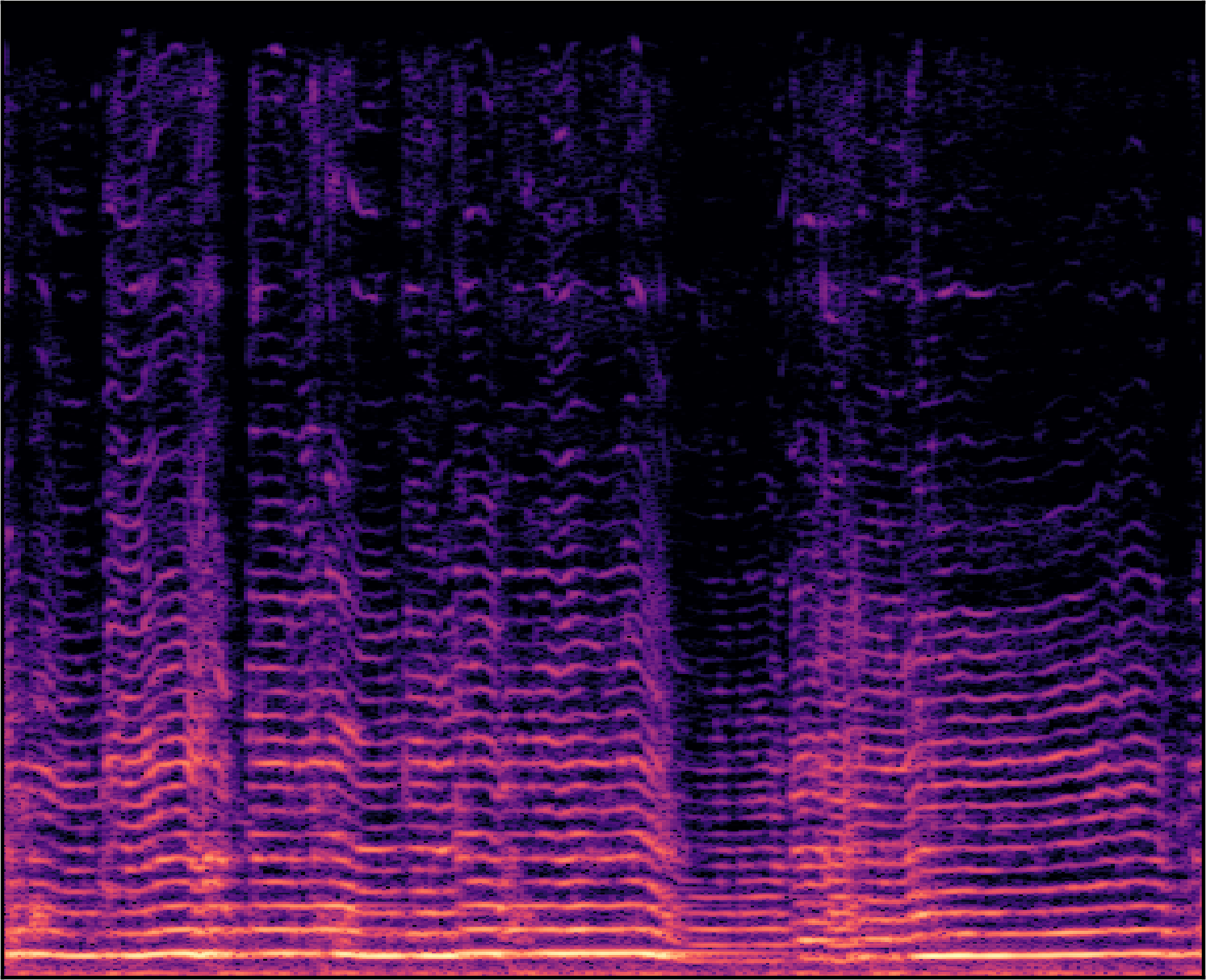}\\
 (a)\\
 \includegraphics[width=.89263\linewidth]{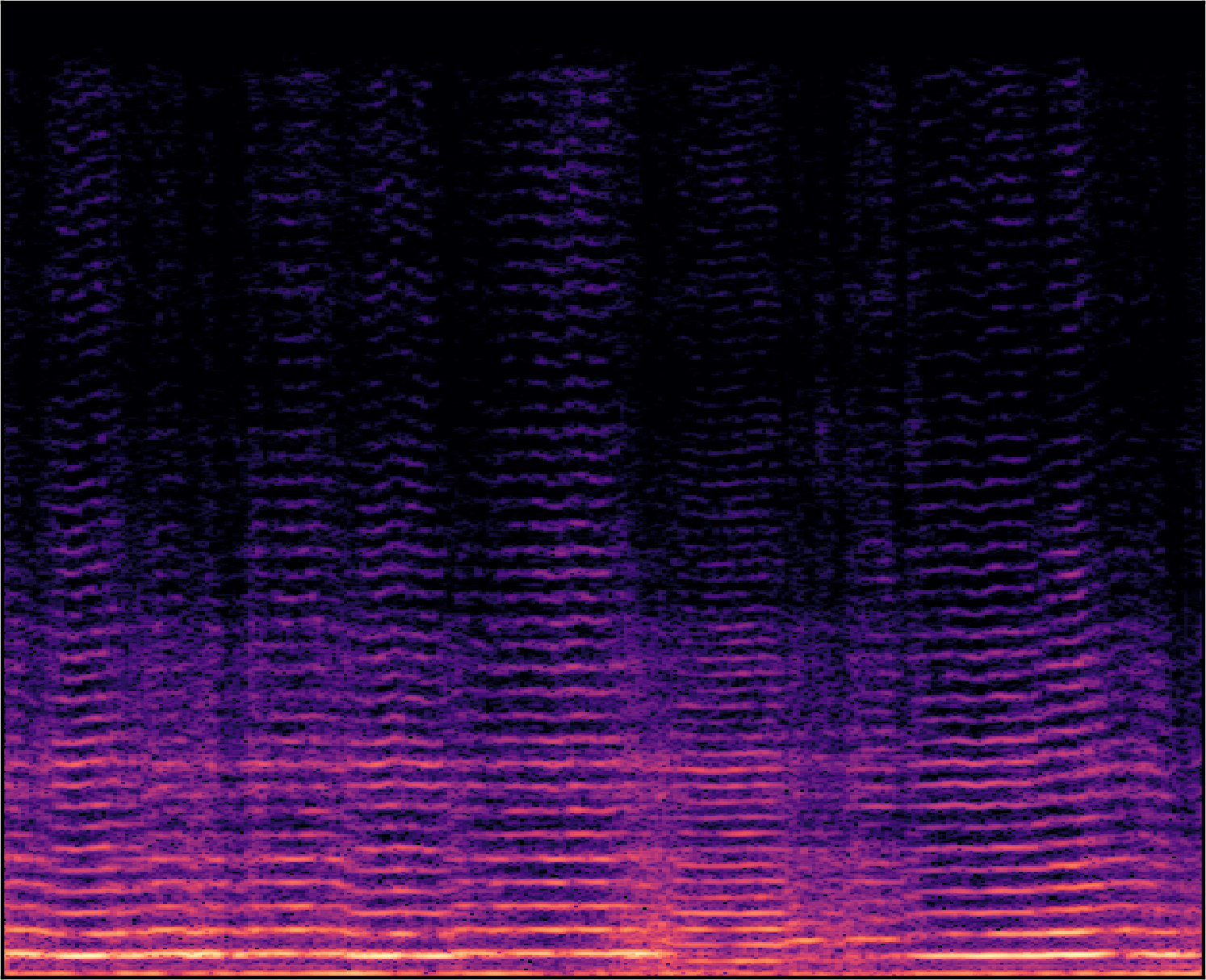}\\
 (b)\\
 \end{tabular}
 \caption{A challenging conversion from a female voice to a cello. (a) The results of DDSP. (b) Our results. While DDSP introduces synthetic noise in order to bridge the different characteristics of the two domains, our method successfully manages to overcome and adapt the input signal to cello's articulation and timbre .}
 \label{fig:compare}
\end{figure}

As can be seen in Tab.~\ref{tab:1}, our method outperforms DDSP both by the melody similarity and target similarity. While the baseline method gets a relatively close score on melody similarity, it is inferior in sound quality and its ability to mimic the target instrument. For example, in some cases  DDSP fails to imitate the target domain timbre, and produces a sine-sounding signal in the correct pitch. An example of a challenging conversion is depicted in Fig.~\ref{fig:compare}.

The high melody preserving results of both methods reflect the fact that both utilize a meaningful f0 sine-wave signal, which aligns the output melody well with the input melody. However, the target similarity results can be explained through the crux of the DDSP mechanism: the learnable function on this network optimizes control parameters of a deterministic noise-additive synthesizer, thus it is upper bounded by the quality of the best-setup synthesizer. Our method, on the other hand, enjoys the expressiveness of a fully capable neural generator, thus can deviate considerably from the source, if needed, in order to generate realistic sounds. 

{\subsection{Data efficiency}}

Another advantage of our method is the need for a minimal amount of training data to generate high quality samples. Successful timbre-transfer results are produced from datasets of few minutes long. For comparison we have trained a state-of-the-art WaveNet based model for music translation~\cite{mor2018autoencoderbased} on two different datasets: URMP, as presented above, and a 30min subset of MusicNet~\cite{musicnet}, as discussed in Sec.~\ref{sec:limits}. In both cases, the music translation method~\cite{mor2018autoencoderbased} failed due to the limited amount of data.

\section{Discussion}\label{sec:discussion}

Recent music AI models vary in the number of parameters and in the required size of the training data. The very recently introduced jukebox model~\cite{jukebox} was trained on over a million songs using hundreds of GPUs, and include 7 billions parameters. The autoencoder-based music-domain translator~\cite{mor2018autoencoderbased} was trained on hours of audio, using tens of GPUs and includes 42M parameters. In comparison, models such as ours are trained on a single GPU, require minutes of audio, and have orders of magnitude less parameters, 1.4M on each scale in our case. The total number of parameters is even smaller than the lean DDSP model, which is of 6M parameters. Taking into account the fact that each scale is trained separately, our model is much more accessible to universities and other small-scale research labs than the other models in the literature.

The method generates sound by shape-shifting a sine-wave, which serves as the skeleton of the rich-timbre painted output. Using scales reflects the inherent structure of the musical audio signal, which is composed of harmonies on different pitch resolutions. The utilization of this strong prior allows us to achieve state-of-the-art results much more efficiently.

The hierarchical structure, which is natural for music generation, also exists in other methods, but in a different way. In the jukebox model, the hierarchy is used separately in the encoders and in the decoders, i.e., all encoder scales are applied, followed by the decoder scales. In our model, there is an interleaving structure in which generation is completed at the lowest scale (including both input encoding and the WaveNet decoder), moving to the processes of the next scale and so on.

\noindent{\bf Limitations\quad}\label{sec:limits}
The economic nature of the model is not without limitations. Unlike the jukebox model, our model does not produce a discrete encoding that can be used (together with an sizable transformer model) for composing new music. In order to add a similar capability, we would need to quantize the input encoding modules ($E^j$) using techniques such as VQ-VAE~\cite{NIPS2017_7210} and to train an auto-regressive model for each level of the hierarchy. Alternatively, any composition method can be used to generate the bare-bones input signal of the network, which would then add the articulation and the timbre to create a richer musical experience.

In the current form, unlike both jukebox and the autoencoder music translator, our model does not share information between different domains, and needs to be retrained on each domain. It is not difficult, however, to modify it to be conditioned on multiple target domains, a path that has been followed many times in the past for other WaveNet-based generators. 

The current method relies on the f0 signal as extracted by a pretrained network that has been trained on monophonic instruments. Since the pitch tracker we employ was trained on monophonic instruments~\cite{crepe}, the results on polyphonic instrument are mostly reasonable but not always.  When successful, our method is successful in transforming the melody to the learned monophonic target domains. However, training polyphonic target instruments remains a challenge since it relies on such a success across the training samples. In the supplementary we present results obtained for polyphonic instruments (keyboard and piano samples from MusicNet), for both our method and DDSP. Both methods succeed to some degree with our method presenting what we consider to be a slight advantage (see supplementary samples). As future work, we note that our method can be readily extended to employ encoders, such as the ones of~\cite{mor2018autoencoderbased,jukebox}, which were trained on large collections of polyphonic music.

\section{Conclusions}\label{sec:conclusions}
We present a novel method of music generation which relies on neural source filtering and hierarchical generation. The method achieves high quality audio generation despite training on small training datasets. The generated input is conditioned on loudness and pitch signals, which are almost source-agnostic, and the characteristic articulation and timbre of the target instrument are introduced through a series of generators. 

\clearpage
\section{Acknowledgments}
We thank Guy Harries and Adam Polyak for helpful discussions. This project has received funding from the European Research Council (ERC) under the European Unions Horizon 2020 research and innovation programme (grant ERC CoG 725974).
\bibliography{music}

\end{document}